\documentclass[prl,aps,twocolumn,showpacs, floatfix]{revtex4}
\usepackage{amssymb}
\usepackage{hyperref}
\usepackage{graphicx}
\usepackage{dcolumn}
\usepackage{bm,amsmath,verbatim}
\usepackage{mathrsfs}

\def\be{\begin{equation}}
\def\ee{\end{equation}}
\def\bea{\begin{eqnarray}}
\def\eea{\end{eqnarray}}
\def\bse{\begin{subequations}}
\def\ese{\end{subequations}}

\def\be{\begin{eqnarray}}
\def\ee{\end{eqnarray}}

\begin{document}

\title{Berezinskii-Kosterlitz-Thouless Phase Transition in 2D Spin-Orbit
Coupled Fulde-Ferrell Superfluids}
\author{Yong Xu}
\author{Chuanwei Zhang}
\thanks{Corresponding Author, Email: chuanwei.zhang@utdallas.edu}
\affiliation{Department of Physics, The University of Texas at Dallas, Richardson, Texas
75080, USA}

\begin{abstract}
The experimental observation of traditional Zeeman-field induced
Fulde-Ferrell-Larkin-Ovchinnikov (FFLO) superfluids has been hindered by
various challenges, in particular, the requirement of low dimensional
systems. In 2D, finite temperature phase fluctuations lead to extremely
small Berezinskii-Kosterlitz-Thouless (BKT) transition temperature for FFLO
superfluids, raising serious concerns regarding their experimental
observability. Recently, it was shown that FFLO superfluids can be realized
using a Rashba spin-orbit coupled Fermi gas subject to Zeeman fields, which
may also support topological excitations such as Majorana fermions in 2D.
Here we address the finite temperature BKT transition issue in this system,
which may exhibit gapped, gapless, topological, and gapless topological FF
phases. We find a large BKT transition temperature due to large effective
superfluid densities, making it possible to observe 2D FF superfluids at
finite temperature. In addition, we show that gapless FF superfluids can be
stable due to their positive superfluid densities. These findings pave the
way for the experimental observation of 2D gapped and gapless FF superfluids
and their associated topological excitations at finite temperature.
\end{abstract}

\maketitle

The exotic Fulde-Ferrell-Larkin-Ovchinnikov (FFLO) superfluids~\cite%
{FuldePR1964,Larkin1964} with finite center-of-mass momentum Cooper pairing
have played a central role in various fields of physics, such as
heavy-fermion superconductors~\cite{HeavySC1,HeavySC2,HeavySC3}, organic
superconductors~\cite{OrganaticSC}, two-dimensional electron gases~\cite%
{Li2011Nature}, and cold Fermi gases~\cite{Hulet2010Nature}. Despite
intensive search, no conclusive evidence of FFLO states has been
experimentally observed. One challenge comes from the narrow region of the
phase diagram where FFLO superfluids exist~\cite{Leo2006PRL}. This region
increases dramatically in lower dimensions~\cite{Mueller2007PRL,Torma2008NJP}%
. However, owing to thermal fluctuations, low dimensional systems cannot
undergo conventional phase transition to a state with long-range order. In
particular, in two dimension (2D), the relevant physics is the
Berezinskii-Kosterlitz-Thouless (BKT) transition~\cite{Bere1971,Thouless} to
a state with quasi-long-range order (\textit{i.e.} vortex-antivortex (V-AV)
pairs) ~\cite{MeloPRL,LianyiHePRL,MingGongPRL}, with the critical
temperature $T_{\mathrm{BKT}}$ determined by the superfluid density tensor.
However, in traditional Zeeman field induced FFLO states, the effective
superfluid density is extremely small (zero for Fulde-Ferrell (FF) states)
due to the rotational symmetry of the Fermi surface, leading to extremely
small $T_{\mathrm{BKT}} $ (zero for FF states)~\cite%
{Vishwanath2009PRL,Torma2014PRB}. Such small $T_{\mathrm{BKT}}$ holds even
for 2D optical lattice systems, although there exists a large parameter
region for the pseudogap phase with the FFLO type of order parameter (but no
phase coherence and superfluidity) \cite{Torma2008NJP,Wolak2012PRA}. This
raises a realistic question whether FFLO states can indeed be observable in
2D at finite temperature.

In the past few years, synthetic spin-orbit (SO) coupling \cite%
{Lin2011Nature,Jing2012PRL,Zwierlen2012PRL,PanJian2012PRL,
Peter2013,Spilman2013PRL, Spilman2013NatRev} has attracted increasing
attention in cold atom community because of its key role in many intriguing
physics such as topological superfluids \cite%
{Zhang2008PRL,Sato2009PRL,ShiLiang2011PRL,LJiang2011PRL,
Lang2012PRL,Yi2013JPB,Xiong2014PRL}, which can accommodate Majorana fermions
in low dimensions with potential applications in fault-tolerant topological
quantum computation~\cite{Kitaev}. In SO coupled Fermi gases with an
in-plane Zeeman field, FF superfluids are dominant in the low temperature
phase diagram due to the asymmetric Fermi surface \cite{Zheng2013PRA,
FanWu2013PRL,Liu2013PRA,Fu2013PRA,LinDongArx,Hui2013PRA, Iskin2013,YongArx13}%
. And with an additional out-of-plane Zeeman field, topological FF
superfluids can emerge~\cite%
{Qu2013NC,Yi2013NC,XJ2013PRA,Chun2013PRL,Yong2014PRL,MGong2014PRB,YGao2014arXiv,Hu2014arXiv}%
. However, previous results are mainly based on mean-field theory, and one
may wonder whether such FF superfluids can be observed experimentally at
finite temperature in 2D through the BKT mechanism. Furthermore, these FF
states exhibit gapless quasiparticle excitations in some parameter regions~%
\cite%
{FanWu2013PRL,LinDongArx,Yong2014PRL,MGong2014PRB,YGao2014arXiv,Hu2014arXiv}%
, and their stability may become a serious issue, similar as the well known
unstable breached pair (BP) phases with \textit{s}-wave contact interactions
due to the divergence of fluctuations~ \cite%
{Vincent2003PRL,Yip2003PRA,Vincent2005PRL}.

In this Letter, we address these crucial issues by studying 2D SO coupled
Fermi gases with both in-plane and out-of-plane Zeeman fields in the
presence of finite temperature phase fluctuations beyond mean-field theory.
Our main findings are that: 1) The finite momenta of Cooper pairs lead to
anisotropic superfluid densities along $x$ and $y$. However,
they are both nonzero and large, in contrast to zero transverse superfluid
density in traditional Zeeman field induced FF superfluids. This leads to a
finite BKT transition temperature, making it possible to observe FF
superfluids in 2D. 2) The superfluid densities for gapless states are
positive, implying that gapless FF superfluids are stable. 3) The changes of
$T_{\mathrm{BKT}}$ with respect to Zeeman fields exhibit an inflection
point, where the gap of the quasiparticle excitation spectrum at zero
momentum closes. In particular, this inflectional behavior is stronger for a
gapless state because of the higher density of states. 4) The anisotropic
superfluid density tensor leads to anisotropic V-AV pairs below $T_{\mathrm{BKT}}$.

Consider a 2D Rashba-type SO coupled Fermi gas with two equal-mass fermion
species, labeled by up and down arrows respectively, subject to both
in-plane ($h_{x}$) and out-of-plane ($h_{z}$) Zeeman fields and $s$-wave
attractive contact interactions. The 2D system can be realized
experimentally by a strong harmonic trap or deep optical lattices, which
freeze atoms to the ground state along the third dimension. The many-body
Hamiltonian reads%
\begin{eqnarray}
H &=&\int d\mathbf{r}\hat{\Psi}^{\dagger }(\mathbf{r})H_{s}(\hat{\mathbf{p}})%
\hat{\Psi}(\mathbf{r}) \\
&&-U\int d\mathbf{r}\hat{\Psi}_{\uparrow }^{\dagger }(\mathbf{r})\hat{\Psi}%
_{\downarrow }^{\dagger }(\mathbf{r})\hat{\Psi}_{\downarrow }(\mathbf{r})%
\hat{\Psi}_{\uparrow }(\mathbf{r}),  \notag
\end{eqnarray}%
where the single particle Hamiltonian $H_{s}(\hat{\mathbf{p}})=\frac{\hat{%
\mathbf{p}}^{2}}{2m}-\mu +H_{\mathrm{SOC}}(\hat{\mathbf{p}})+H_{z}$ with
momentum operator $\hat{\mathbf{p}}=-i\hbar (\partial _{x}\mathbf{e}%
_{x}+\partial _{y}\mathbf{e}_{y})$, chemical potential $\mu $, attractive
interaction strength $U$, and atom mass $m$. The Rashba SO coupling $H_{%
\mathrm{SOC}}(\hat{\mathbf{p}})=\alpha ({\bm\sigma }\times \hat{\mathbf{p}}%
)\cdot {\mathbf{e}_{z}}$ with Pauli matrices $\mathbf{\sigma }$; the Zeeman
field $H_{z}=h_{x}\sigma _{x}+h_{z}\sigma _{z}$ along $x$ and $z$. $\hat{\Psi%
}(\mathbf{r})=[\hat{\Psi}_{\uparrow }(\mathbf{r}),\hat{\Psi}_{\downarrow }(%
\mathbf{r})]^{T}$ and $\hat{\Psi}_{\nu }^{\dagger }(\mathbf{r})$ ($\hat{\Psi}%
_{\nu }(\mathbf{r})$) creates (annihilates) a fermionic atom at $\mathbf{r}$.

In quantum field theory, the partition function at temperature $T=1/\beta $
can be written as a path integral (See supplementary materials S-1) $Z=\int
D(\bar{\Delta},\Delta )e^{-S_{\mathrm{eff}}[\bar{\Delta},\Delta ]}$ with the
effective action written as $S_{\mathrm{eff}}[\bar{\Delta},\Delta
]=\int_{0}^{\beta }d\tau \int d\mathbf{r}\frac{|\Delta |^{2}}{U}-\frac{1}{2}%
\ln \det {G}^{-1},$ where the inverse single particle Green function $%
G^{-1}=-\partial _{\tau }-H_{\mathrm{B}}$ in the Nambu-Gor$^{,}$kov basis,
with $4\times 4$ Bogoliubov-de Gennes (BdG) Hamiltonian%
\begin{equation}
H_{\mathrm{B}}=\left(
\begin{array}{cc}
H_{s}(\hat{\mathbf{p}}) & \Delta (\mathbf{r},\tau ) \\
\Delta (\mathbf{r},\tau ) & -\sigma _{y}H_{s}(\hat{\mathbf{p}})^{\ast
}\sigma _{y}%
\end{array}%
\right) .
\end{equation}%
By assuming $\Delta (\mathbf{r})=\Delta _{0}e^{iQ_{y}y}$ given the Fermi
surface asymmetry along \textit{y}~\cite{YongArx13}, the mean-field
saddle point without phase fluctuations can be obtained by the saddle
equations $\partial \Omega /\partial \Delta _{0}=0$, $\partial \Omega
/\partial Q_{y}=0$, and the particle number equation $\partial \Omega
/\partial \mu =-n$ with a fixed total density $n$~\cite{Yong2014PRL}. $%
\Omega =S_{\mathrm{eff}}/\beta $ is the thermodynamical potential. Here the
ultra-violet divergence can be regularized by $1/U=\sum_{\mathbf{k}}1/(\hbar
^{2}k^{2}/m+E_{b})$ with the binding energy $E_{b}$.

To study the effects of phase fluctuations, we set $\Delta (\mathbf{r},\tau
)=\Delta _{0}e^{iQ_{y}y+i\theta (\mathbf{r},\tau )}$, where $\theta (\mathbf{%
r},\tau )$ is the phase fluctuations around the saddle point. Note that we
have neglected the amplitude fluctuations corresponding to the gapped
excitations. The inverse Green function $G^{-1}=G_{0}^{-1}+\Sigma $ with the
mean-field one $G_{0}^{-1}$ and self-energy $\Sigma =(i\partial _{\tau
}\theta /2+\hbar ^{2}Q_{y}\partial _{y}\theta /(4m)+\hbar ^{2}(\nabla \theta
)^{2}/(8m))\sigma _{z}\otimes \sigma _{0}+\hbar \nabla \theta \cdot \hat{%
\mathbf{p}}/(2m)-i\hbar ^{2}\nabla ^{2}\theta /(4m)+(\alpha \hbar \partial
_{x}\theta /2)(\sigma _{0}\otimes \sigma _{y})-(\alpha \hbar \partial
_{y}\theta /2)(\sigma _{0}\otimes \sigma _{x})$. The effective action $S_{%
\mathrm{eff}}=S_{\mathrm{eff}}^{0}+S_{\mathrm{eff}}^{\mathrm{fluc}}$ with
the contribution of the fluctuations $S_{\mathrm{eff}}^{\mathrm{fluc}}=\text{%
tr}\sum_{l=1}^{\infty }(G_{0}\Sigma )^{l}/(2l)$. Expanding it to the second
order yields
\begin{eqnarray}
S_{\mathrm{eff}}^{\mathrm{fluc}} =&&\frac{1}{2}\int_{0}^{\beta }d\tau \int d{%
\mathbf{r}}\left[ J_{xx}(\partial _{x}\theta )^{2}+J_{yy}(\partial
_{y}\theta )^{2}\right.  \notag \\
&&\left.+J_{\tau y}i\partial _{\tau }\theta \partial _{y}\theta +P(\partial
_{\tau }\theta )^{2}-iA\partial _{\tau }\theta \right] ,
\end{eqnarray}
with the superfluid density tensor $J_{\nu \mu }$ (here $J_{xy}=0$), the
corresponding superfluid density $\rho _{\mu \nu }=4mJ_{\mu \nu}/(\hbar
^{2}n) $ scaled by the fixed total density, and the compressibility $P$.
Compared with the formula for the normal superfluids~\cite{MingGongPRL}, one
additional term $J_{\tau y}$ emerges because of the nonzero Cooper pairing
momenta $Q_{y}$. Note that these parameters cannot be expressed analytically
and they are obtained via numerical approach (See supplementary materials
S-1). We have checked that the results without $h_{x}$ are exactly the same
as previous ones~\cite{MingGongPRL}.

By decomposing the phase $\theta (\mathbf{r},\tau )$ into two parts: a
static vortex configuration $\theta (\mathbf{r})_{\mathrm{v}}$ and a
time-dependent spin-wave one $\theta (\mathbf{r},\tau )_{\mathrm{sw}}$, the
effective action contributed by phase fluctuations can be written as $S_{%
\mathrm{eff}}^{\mathrm{fluc}}=S_{\mathrm{eff}}^{\mathrm{v}}+S_{\mathrm{eff}%
}^{\mathrm{sw}}$, where the vortex part $S_{\mathrm{eff}}^{\mathrm{v}}=\frac{%
1}{2}\int d{r}\sum_{\nu =x,y}J_{\nu \nu }(\partial _{\nu }\theta _{\mathrm{v}%
})^{2}$ and spin-wave part $S_{\mathrm{eff}}^{\mathrm{sw}}=\frac{1}{2}\int d{%
r}[\sum_{\nu =x,y}J_{\nu \nu }(\partial _{\nu }\theta _{\mathrm{sw}%
})^{2}+J_{\tau y}i\partial _{\tau }\theta \partial _{y}\theta
_{sw}+P(\partial _{\tau }\theta _{\mathrm{sw}})^{2}-iA\partial _{\tau
}\theta _{\mathrm{sw}}]$. The integration of the spin-wave part gives $S_{%
\mathrm{eff}}^{\mathrm{sw}}=\sum_{\mathbf{k}}\mathrm{{ln}(1-e^{-\beta E_{sw}(%
\mathbf{k})})}$ where the spin-wave excitation $E_{\mathrm{sw}}(\mathbf{k}%
)=\frac{1}{2P}[-J_{\tau y}k_{y}+\sqrt{J_{\tau
y}^{2}k_{y}^{2}+4P(J_{xx}k_{x}^{2}+J_{yy}k_{y}^{2}))}]$. This
anisotropic linear spectrum has anisotropic sound speeds: $v_{x}=\sqrt{%
J_{xx}/P}$ along \textit{x} and $v_{y\pm}=\frac{1}{2P}(\mp J_{\tau y}+\sqrt{J_{\tau y}^{2}+4PJ_{yy}})$
along positive and negative \textit{y} directions~\cite{Yong2014PRL}.
We note that the anisotropic sound speed also exists in other anisotropic
superfluid systems~\cite{Melo2014arXiv}, whereas the anisotropic behavior
between the opposite directions among superfluids can occur only in FF
superfluids. In normal BCS superfluids, $Q_{y}=0$ and $v=\sqrt{J_{xx}/P}$~%
\cite{MeloPRL,MingGongPRL}.

With phase fluctuations, the parameters $\Delta _{0}$, $Q_{y}$, and $\mu $
can be calculated by self-consistently solving the saddle equations $%
\partial \Omega ^{0}/\partial \Delta _{0}=0$, $\partial \Omega /\partial
Q_{y}=0$, and the particle number equation $\partial \Omega /\partial \mu =-n
$, where $\Omega ^{0}=S_{\mathrm{eff}}^{0}/\beta $ and $\Omega =S_{\mathrm{%
eff}}/\beta $. Instead of the critical temperature determined by $\Delta
_{0}=0$ in the mean-field theory, the critical BKT temperature is determined
\cite{Thouless,Torma2014PRB,Melo2014arXiv} (See supplementary materials S-2)
by
\begin{equation}
T_{\mathrm{BKT}}=\frac{\pi }{2}\sqrt{\prod\nolimits_{\nu =x,y}J_{\nu \nu
}(\Delta _{0},Q_{y},\mu ,T_{\mathrm{BKT}})}.  \label{TBKT}
\end{equation}%
Note that $J_{xx}$ and $J_{yy}$ can be renormalized by the renormalization
group theory, which does not change the physics qualitatively~\cite%
{StoofColdField}. Across $T_{\mathrm{BKT}}$ that is much lower than the
mean-field critical temperature, Fermi gases transit from a pseudogap phase
(with nonzero $\Delta _{0}$ but without phase coherence) to a superfluid
affluent with V-AV pairs (with both pairing and phase coherence). We
calculate $T_{\mathrm{BKT}}$ by solving the saddle point equations, the
particle number equation, and Eq.(\ref{TBKT}) self-consistently. Here, the
energy unit is chosen as the Fermi energy $E_{\mathrm{F}}=\hbar ^{2}{K}_{%
\mathrm{F}}^{2}/2m$ with Fermi vector $K_{\mathrm{F}}=(2\pi n)^{1/2}$.

\begin{figure}[t]
\includegraphics[width=3.4in]{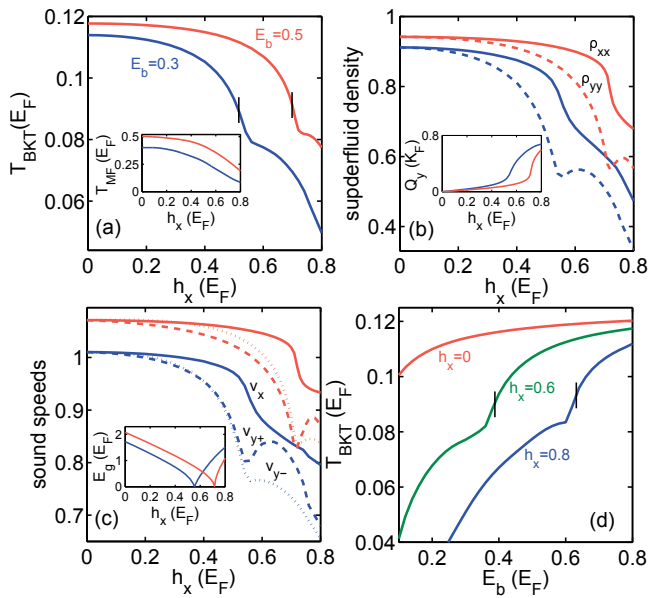}
\caption{(Color online) Plot of $T_{\mathrm{BKT}}$ (in (a)), superfluid
densities and sound speeds (in (b) and (c)) evaluated at $T_{\mathrm{BKT}}$,
as a function of $h_{x} $ with $E_{b}=0.3E_{\mathrm{F}}$ and $E_{b}=0.5E_{%
\mathrm{F}}$. The transition from gapped to gapless superfluids is marked by
a short vertical line. In the inset of (a), (b), and (c), the mean-field
critical temperatures, the momentum of Cooper pairs $Q_{y}$ at $T_{\mathrm{%
BKT}}$, and the gap at zero momentum at $T_{\mathrm{BKT}}$ are plotted
respectively. The unit of the speed of sounds is $v_{\mathrm{F}}/\protect%
\sqrt{2}$ with Fermi velocity $v_{\mathrm{F}}$ and the unit of superfluid
densities is $n$. In (b), the solid and dashed lines correspond to $\protect%
\rho _{xx}$ and $\protect\rho _{yy}$ respectively. In (c), the solid,
dashed, and dotted lines respectively correspond to the sound speeds along
the \textit{x} direction $v_{x}$, positive \textit{y} direction $v_{y+} $,
and negative \textit{y} direction $v_{y-}$. (d) Plot of $T_{\mathrm{BKT}}$
with respect to the binding energy $E_{b}$ at fixed $h_{x}$. Here $\protect%
\alpha K_{\mathrm{F}}=E_{\mathrm{F}}$ and $h_{z}=0$.}
\label{Gapless_FF}
\end{figure}

The BKT phase transition for FF states was studied for imbalanced Fermi
gases without SO coupling and it was found that the superfluid density in
the direction perpendicular to the finite momenta of Cooper pairs is zero
due to the rotational invariance of the Fermi surface~\cite{EZeroBKT},
suggesting that FF superfluids may not be observable at finite temperature
in 2D. Even considering the LO state, the critical temperature is still much
lower because of the extremely high anisotropy of the superfluid density~%
\cite{Vishwanath2009PRL}. However, in Fig.~\ref{Gapless_FF}, we find that
the BKT critical temperature $T_{\mathrm{BKT}}$ for the FF superfluids is
finite and large. Although the superfluid densities are still anisotropic
with $\rho _{xx}>\rho _{yy}$ (shown in Fig.~\ref{Gapless_FF} (b)) due to the
deformation of the equal thermodynamic potential along \textit{y}, this
anisotropy is not high enough to destroy the superfluidity at finite
temperature in sharp contrast to $\rho _{yy}>\rho _{xx}=0$ for traditional
FF states without SO coupling~\cite{Vishwanath2009PRL,Torma2014PRB}. This
result provides the theoretical foundation for the feasibility of observing
FF states at finite temperature in 2D systems. Furthermore, $T_{\mathrm{BKT}}
$ is much lower than mean-field transition temperature (shown in the inset
of Fig.~\ref{Gapless_FF} (a)) as expected, whereas the finite momenta (shown
in the inset of Fig.~\ref{Gapless_FF} (b)) evaluated at $T_{\mathrm{BKT}}$)
of Cooper pairs are not destroyed by phase fluctuations.


\begin{figure}[t]
\includegraphics[width=3.4in]{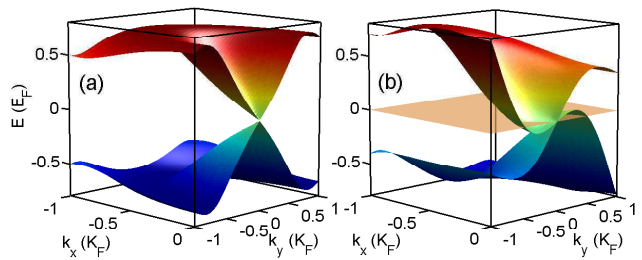}
\caption{(Color online) Quasiparticle excitations in the $(k_{x},k_{y})$
plane at the critical point (a) from normal superfluids to topological
superfluids, where $h_{x}=0$ and $h_{z}=0.652E_{\mathrm{F}}$; and (b) from
gapless FF superfluids to gapless topological FF superfluids, where $%
h_{x}=0.5E_{\mathrm{F}}$ and $h_{z}=0.282E_{\mathrm{F}}$. The physical
quantities are evaluated at $T_{\mathrm{BKT}}$. The light color plane in (b)
corresponds to zero excitation energy. Here $\protect\alpha K_{\mathrm{F}%
}=E_{\mathrm{F}}$ and $E_{b}=0.3E_{\mathrm{F}}$.}
\label{excitation}
\end{figure}

With increasing $h_{x}$, the FF superfluids transit from gapped to gapless
states \cite%
{FanWu2013PRL,LinDongArx,Yong2014PRL,MGong2014PRB,YGao2014arXiv,Hu2014arXiv}
due to the strong distortion of quasiparticle excitations along \textit{y}.
The gapless FF superfluids exhibit a Fermi surface in quasiparticle spectrum
as shown in Fig.~\ref{excitation}(b). For a gapless superfluid, it is
important to inquire whether they are stable. For instance, the famous
gapless BP phase~\cite{Vincent2003PRL} with simple $s$-wave contact
interactions was shown to be unstable with a negative superfluid density~%
\cite{Yip2003PRA, Vincent2005PRL}. In Fig.~\ref{Gapless_FF}, we see that the
superfluid densities for gapless states (the critical point of which is
marked by a short line) are positive and $T_{\mathrm{BKT}}$ is finite,
implying that the superfluids are stable. Furthermore, there is an
inflection point near the gapless transition point, corresponding to the
minimum of the superfluid density and sound speeds. This point is exactly
where the gap at zero momentum of quasiparticle excitations (shown in the
inset of Fig.~\ref{Gapless_FF}(c)) closes as ${\bar{h}_{x}}^{2}={\bar{\mu}}%
^{2}+\Delta _{0}^{2}$ with $\bar{h}_{x}=h_{x}+\alpha Q_{y}/2$ and $\bar{\mu}%
=\mu -Q_{y}^{2}/8m$, suggesting that phase fluctuations have dramatic
effects at zero momentum (long wavelength limit), whereas the gapless
surface emerges at nonzero momenta. Analogous to the anisotropic superfluid
density tensor, the sound speeds shown in Fig.~\ref{Gapless_FF}(c) are
anisotropic along the \textit{x} and \textit{y} directions and even
different along the positive and negative \textit{y} directions, which is
unique for FF superfluids. The anisotropy increases with the increasing of $%
Q_{y}$ with respect to $h_{x}$ until the speed of sounds reaches the minimum
where the gap at zero momentum closes. At $h_{x}=0$, $v_{x}=v_{y+}=v_{y-}$
with the value close to $v_{\mathrm{F}}/\sqrt{2}$ in the BCS limit.

In Fig.~\ref{Gapless_FF} (d), we plot the BKT temperature with respect to
the binding energy $E_{b}$ at fixed Zeeman fields. $T_{\mathrm{BKT}}$ is a
monotonically increasing function of $E_{b}$, and approaches a constant
value at large $E_{b}$ that is independent of $h_{x}$, signaling the
crossover from BCS Cooper pairs to Bose-Einstein condensates of tightly
bound molecules. Also, the increased $\Delta_0$ by increasing $E_b$ closes
the gap at zero momentum when ${\bar{h}_x}^2={\bar\mu}^2+\Delta_0^2$, where
there emerges an inflection point accompanied by the critical transition
point between gapless and gapped FF superfluids. Without Zeeman fields ($%
h_x=0$), no gap closing point at zero momentum appears and thus no
inflection point.

\begin{figure}[t]
\includegraphics[width=2.6in]{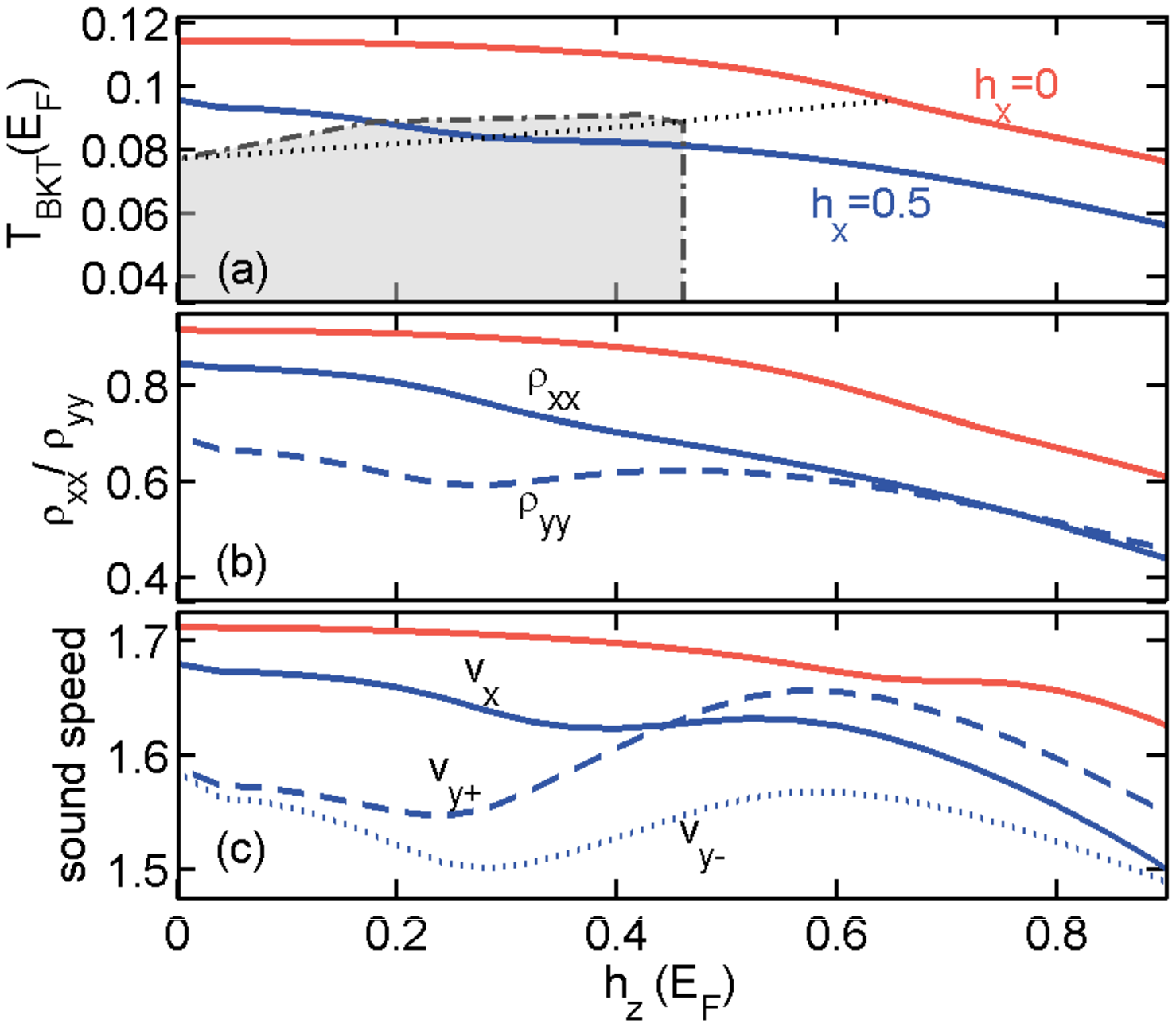}
\caption{(Color online) Plot of $T_{\mathrm{BKT}}$ (in (a)), superfluid
density (in (b)) and sound speeds (in (c)) evaluated at $T_{\mathrm{BKT}}$,
as a function of $h_{z}$ with $h_{x}$ fixed. The states are FF superfluids
except the one with $h_{x}=0$. The black dotted line connects the
topological transition points as $h_{x}$ varies. The grey area surrounded by
the dashed-dotted line maps out the gapless state region. The red and blue
lines correspond to $h_{x}=0,0.5E_{\mathrm{F}}$ respectively in all figures.
Here $\protect\alpha K_{\mathrm{F}}=E_{\mathrm{F}}$ and $E_{b}=0.3E_{\mathrm{F}}$.}
\label{TS_FF}
\end{figure}

To demonstrate the effects of $h_{z}$ Zeeman field, we plot the change of
the BKT temperature $T_{\mathrm{BKT}}$ ((a)), the superfluid density ((b)),
and the speeds of sound ((c)) with respect to $h_{z}$ for fixed $h_{x}$ in
Fig.~\ref{TS_FF}. Clearly, $T_{\mathrm{BKT}}$ is a monotonically decreasing
function of Zeeman fields since both Zeeman fields are detrimental to the
Cooper pairing. Superfluid density and sound speeds are both anisotropic
similar to the pure $h_{x}$ scenario. On the other hand, given that the
quasiparticle gap at zero momentum first closes and then reopens at the
topological transition point~\cite%
{Qu2013NC,Yi2013NC,XJ2013PRA,Chun2013PRL,Yong2014PRL,MGong2014PRB,YGao2014arXiv,Hu2014arXiv}
$h_{z}^{2}={\bar{\mu}}^{2}+\Delta _{0}^{2}-{\bar{h}_{x}}^{2}$ driven by $%
h_{z}$ field, one may expect an inflectional or minimum behavior at this
point, similar to the case with pure $h_{x}$ field. Indeed, there is a
manifest infection point for $T_{\mathrm{BKT}}$ and minimum points for
superfluid density and sound speeds in gapless topological superfluids (e.g.
blue line in Fig.~\ref{TS_FF}). However, this inflectional or minimum
behavior is much smaller in either topological superfluids or topological FF
superfluids with zero or small $h_{x}$ (e.g. red line in Fig.~\ref{TS_FF}.
See also supplementary materials S-3) because of the lower density of states
at zero momentum compared with that in gapless topological superfluids as
visualized in Fig.~\ref{excitation}.

In Fig.~\ref{TS_FF}(a), the dotted line represents $T_{\mathrm{BKT}}$ at the
topological transition points as $h_{x}$ varies, showing that $T_{\mathrm{BKT%
}}$ decreases with increasing $h_{x}$ while the critical $h_{z}$ decreases~%
\cite{Qu2013NC}. This indicates that the existence of $h_{x}$ cannot enhance
the BKT temperature for the observation of topological superfluids. The grey
region~\cite{Egapless} surrounded by the dashed-dotted line is where the
gapless superfluids can be observed. Compared with the 3D case where the
gapless superfluids (including both topological and topological trivial
phases) are dominant~\cite{Yong2014PRL}, this gapless region is much smaller
because the gap at $\mathbf{k}=0$ reopens across the topological transition
point.

Below the BKT temperature, the superfluids are affluent with V-AV pairs,
which can be obtained by solving $\nabla \times \mathbf{v}_{s}=2\pi
\sum_{i}n_{i}\delta (\mathbf{r}-\mathbf{r}_{i})$ and the minimization of $S_{%
\mathrm{eff}}^{\mathrm{v}}$: $\rho _{xx}\partial _{x}^{2}\theta _{v}+\rho
_{yy}\partial _{y}^{2}\theta _{v}=0$. Here $\mathbf{v}_{s}=\nabla \theta _{%
\mathrm{v}}(\mathbf{r})$, $n_{i}=\pm 1$ represents a vortex or antivortex
localized at $\mathbf{r}_{i}$. For a single vortex, $\theta _{\mathrm{v}}=%
\text{arctan}(\rho ^{2}y/x)$ with $\rho =(\rho _{xx}/\rho _{yy})^{1/4}$. For
a V-AV pair located at $(\pm x_{0},0)$, $\theta _{\mathrm{v}}=\arctan [2\bar{%
x_{0}}\bar{y}/(\bar{x_{0}}^{2}-\bar{x}^{2}-\bar{y}^{2})]$ with $\bar{x}%
=x/\rho $,$\bar{y}=y\rho $, and $\bar{a}=a/\rho $. Interestingly, the
anisotropic superfluid densities induced by the finite momentum pairing
results in anisotropic V-AV pairs (See supplementary materials S-4) in
contrast to isotropic ones in normal BCS superfluids for Rashba-type SO
coupled Fermi gases~\cite{MeloPRL,ZhiDong2012PRL,LianyiHePRL,MingGongPRL}.
Anisotropic V-AV pairs have also been discussed recently where the
anisotropy is caused by the anisotropic SO coupling~\cite{Melo2014arXiv},
instead of FF pairing here.

For Fermi gases, the pseudogap phenomenon beyond the BKT temperature has
been observed in 2D $^{40}$K fermionic atoms with the number of atoms on the
order of $10^{3}$ and the harmonic trap frequency $2\pi \times 127$Hz~\cite%
{Feld2011Natrue}. Equal Rashba and Dresselhaus (ERD) SO coupling as well as
Zeeman fields have also been engineered in these atom gases~\cite%
{Jing2012PRL,Zwierlen2012PRL} by coupling two hyperfine states using Raman
lasers~\cite%
{Lin2011Nature,Jing2012PRL,Zwierlen2012PRL,PanJian2012PRL,Peter2013,Spilman2013PRL, Spilman2013NatRev}%
. The experimental realization of Rashba SO coupling is currently under
investigation. The large BKT temperature conclusion should also apply to the
FF states in the ERD-type SO coupled Fermi superfluids because of the same
mechanism for inducing FF type Cooper pairs~\cite{FanWu2013PRL,YongArx13}.
Recently, the observation of quasi long-range order in 2D $^{6}$Li gases has
been reported~\cite{Jochim2014arXiv} and the number of atoms can be as large
as 5$\times 10^{4}$. Similar experimental setup may be employed to realize
the FF superfluids in 2D with the feature: anisotropic sound speeds, which
can be experimentally probed through the density perturbation~\cite%
{Ketterle1997PRL}. Also other methods, such as observing the anisotropic
V-AV pairs by time-of-flight expansion or Bragg scattering can be
considered. We note that in real experiments the BKT transition occurs as
the BKT crossover because of finite-size effects as shown in the 2D BEC
experiment~\cite{Hadzibabic2006Nature}.

In summary, we investigate the BKT phase transition in a 2D SO coupled Fermi
gas subject to Zeeman fields and find finite BKT temperatures for both
gapped and gapless FF superfluids in sharp contrast to the case without SO
coupling, where it is zero due to the vanishing transverse superfluid
density. Our findings demonstrate the feasibility for the experimental
observation of FF superfluids (gapped or gapless, topological or
non-topological) and the associated topological excitations (e.g., Majorana
fermions) in a 2D SO coupled Fermi gas at finite temperature.

\begin{acknowledgments}
\textbf{Acknowledgements}: This work is supported by ARO (W911NF-12-1-0334)
and AFOSR (FA9550-11-1-0313 and FA9550-13-1-0045). We thank Texas Advanced
Computing Center (TACC), where our numerical simulations were performed.
\end{acknowledgments}

\begin{widetext}
\maketitle
\setcounter{equation}{0} \setcounter{figure}{0} \setcounter{table}{0} %
\renewcommand{\theequation}{S\arabic{equation}} \renewcommand{\thefigure}{S%
\arabic{figure}} \renewcommand{\bibnumfmt}[1]{[S#1]} \renewcommand{%
\citenumfont}[1]{S#1}

In the main text we present various physical quantities
including the BKT temperature, superfluid density tensor, and speed of sound
for a spin-orbit coupled Fermi gas subject to both in-plane and out-of-plane
Zeeman fields. Here we provide more detailed calculation information in Sec.
S-1 and S-2, plot the BKT temperature, superfluid density, and sound speeds
for more parameters in Sec. S-3, and also plot anisotropic V-AV pair structures in Sec. S-4.

\section{S-1. DERIVATION OF SUPERFLUID DENSITY TENSOR AND SOUND SPEED}

In quantum field theory, the partition function can be written as $Z=\text{Tr%
}(e^{-\beta H})=\int D(\bar{\psi},\psi )e^{-S_{\rm eff}[\bar{\psi},\psi ]}$ with
$\beta =1/T$ at the temperature $T$. The effective action is
\begin{equation}
S_{\rm eff}[\bar{\psi},\psi ]=\int_{0}^{\beta }d\tau \left( \int d\mathbf{r}%
\sum_{\sigma }\bar{\psi}_{\sigma }(\mathbf{r},\tau )\partial _{\tau }\psi
_{\sigma }(\mathbf{r},\tau )+H(\bar{\psi},\psi )\right) ,
\end{equation}%
where $\int d\tau $ is an integral over the imaginary time $\tau $ and $H(%
\bar{\psi},\psi )$ is obtained by replacing $\hat{\Psi}_{\sigma }^{\dagger }$
and $\hat{\Psi}_{\sigma }$ with Grassman field number $\bar{\psi}_{\sigma }$
and $\psi _{\sigma }$. We can transform the quartic interaction term to
quadratic one by Hubbard-Stratonovich transformation, where the order
parameter $\Delta (\mathbf{r},\tau )$ is defined. By integrating out fermion
fields, the partition function becomes $Z=\int D(\bar{\Delta},\Delta
)e^{-S_{\rm eff}[\bar{\Delta},\Delta ]}$, where the effective action can be
written as
\begin{equation}
S_{\rm eff}[\bar{\Delta},\Delta ]=\int_{0}^{\beta }d\tau \int d\mathbf{r}(\frac{%
|\Delta |^{2}}{U})-\frac{1}{2}\ln \det {G}^{-1}.  \label{Seff}
\end{equation}%
Here the inverse single particle Green function $G^{-1}=-\partial _{\tau
}-H_{\rm B}$ in the Nambu-Gor'kov representation with $4\times 4$ Bogoliubov-de
Gennes (BdG) Hamiltonian (Eq.(2) in the main text)
\begin{equation}
H_{\rm B}=\left(
\begin{array}{cc}
H_{s}(\hat{\mathbf{p}}) & \Delta (\mathbf{r},\tau ) \\
\Delta (\mathbf{r},\tau ) & -\sigma _{y}H_{s}(\hat{\mathbf{p}})^{\ast
}\sigma _{y}%
\end{array}%
\right) .
\end{equation}

Assume that a mean-field solution has the FF form $\Delta (\mathbf{r},\tau
)_{0}=e^{iQ_{y}y}\Delta _{0}$ with the space independent $\Delta _{0}$ given
that the in-plane Zeeman field deforms the Fermi surface along the \textit{y}
direction, leading to finite momentum pairing along that direction~\cite%
{YongArx13S}. Through Fourier transformation and the summation of Matsubara
frequency, this form of $\Delta (\mathbf{r},\tau )$ yields mean-field
thermodynamical potential~\cite{Zheng2013PRAS,Yong2014PRLS}.

To study the effects of phase fluctutations, we assume $\Delta (\mathbf{r}%
)=\Delta _{0}e^{iQ_{y}y+i\theta (\tau ,\mathbf{r})}$ with phase fluctuation
field $\theta (\tau ,\mathbf{r})$ around the saddle point. The unitary
transformed inverse Green function becomes
\begin{equation}
G^{-1}=G_{0}^{-1}+\Sigma ,  \label{Green}
\end{equation}%
via the unitary operator
\begin{equation}
U=\left(
\begin{array}{cc}
e^{i(\theta +Q_{y}y)} & 0 \\
0 & e^{-i(\theta +Q_{y}y)}%
\end{array}%
\right) .
\end{equation}%
Here $G_{0}^{-1}$ represents the mean-field part and
\begin{eqnarray}
\Sigma &=&\left[ i\partial _{\tau }\theta /2+\hbar ^{2}Q_{y}\partial
_{y}\theta /(4m)+\hbar ^{2}(\nabla \theta )^{2}/(8m)\right] \sigma _{z}\otimes
\sigma _{0}+\hbar \nabla \theta \cdot \hat{\mathbf{p}}/(2m) \\
&&-i\hbar ^{2}\nabla ^{2}\theta /(4m)+(\alpha \hbar \partial _{x}\theta
/2)(\sigma _{0}\otimes \sigma _{y})-(\alpha \hbar \partial _{y}\theta
/2)(\sigma _{0}\otimes \sigma _{x}) ,  \notag
\end{eqnarray}%
is the self-energy contributed by phase fluctuations. Substituting Eq.~\ref%
{Green} to Eq.~\ref{Seff} leads to the effective action
\begin{equation}
S_{\rm eff}=S_{\rm eff}^{0}+S_{\rm eff}^{\rm fluc},
\end{equation}%
where $S_{\rm eff}^{\rm fluc}=\text{tr}\sum_{l=1}^{\infty }(G_{0}\Sigma )^{l}/(2l)$
represents phase fluctuation contributions. Expanding it to the second order
yields (Eq.(3) in the main text)
\begin{eqnarray}
S_{\rm eff}^{\rm fluc} &=&\frac{1}{2}\text{tr}{G}_{0}\Sigma +\frac{1}{4}\text{tr}%
({G}_{0}\Sigma {G}_{0}\Sigma ) \\
&=&\frac{1}{2}\int d\mathbf{r}\int d\tau \left[ J_{xx}(\partial _{x}\theta
)^{2}+J_{yy}(\partial _{y}\theta )^{2}+J_{xy}\partial _{x}\theta \partial
_{y}\theta +iJ_{\tau y}\partial _{\tau }\theta \partial _{y}\theta +iJ_{\tau
x}\partial _{\tau }\theta \partial _{x}\theta +P(\partial _{\tau }\theta
)^{2}-iA\partial _{\tau }\theta \right] ,  \notag
\end{eqnarray}%
where 

\begin{eqnarray}
J_{xx} & =&\frac{\hbar^{2}}{4m}n+\frac{1}{8\beta (2\pi)^2}\int d\mathbf{k}%
\sum_{\omega_n} \left(\hbar^{2}\alpha^{2}f_{44}+\frac{\hbar^{2}}{m^{2}}%
f_{22}k_{x}^{2}+ \frac{2\hbar^{2}\alpha}{m}f_{24}k_{x}\right), \\
J_{yy} & =&\frac{\hbar^{2}}{4m}n+\frac{1}{8\beta (2\pi)^2}\int d\mathbf{k}
\sum_{\omega_n} \left(\frac{\hbar^{4}}{4m^{2}}f_{11}Q_{y}^{2}-\frac{%
\hbar^{3}\alpha}{m}f_{15}Q_{y}+ \hbar^{2}\alpha^{2}f_{55}+\frac{\hbar^{3}}{%
m^{2}}f_{12}Q_{y}k_{y}+ \frac{\hbar^{2}}{m^{2}}f_{22}k_{y}^{2}-\frac{%
2\hbar^{2}\alpha}{m}f_{25}k_{y}\right), \\
J_{xy} & =&\frac{1}{4\beta (2\pi)^2}\int d\mathbf{k}\sum_{\omega_n}\left(
\frac{\hbar^{3}\alpha}{2m}f_{14}Q_{y}-\hbar^{2}\alpha^{2}f_{45}+ \frac{%
\hbar^{3}}{2m^{2}}f_{12}Q_{y}k_{x}+\frac{\hbar^{2}}{m^{2}}f_{22}k_{x}k_{y}+
\frac{\hbar^{2}\alpha}{m}f_{24}k_{y}-\frac{\hbar^{2}\alpha}{m}%
f_{25}k_{x}\right), \\
J_{\tau y} & =&\frac{1}{8\beta (2\pi)^2} \int d\mathbf{k}\sum_{\omega_n}
\left(\frac{\hbar^{2}}{m}f_{11}Q_{y}-2\hbar\alpha f_{15}+\frac{2\hbar}{m}%
f_{12}k_{y}\right), \\
J_{\tau x} & =&\frac{1}{4\beta (2\pi)^2}\int d\mathbf{k} \sum_{\omega_n}
\left(\hbar\alpha f_{14}+\frac{\hbar}{m}f_{12}k_{x}\right), \\
P & =&-\frac{1}{8}\frac{1}{\beta (2\pi)^2}\int d\mathbf{k}%
\sum_{\omega_n}f_{11}, \\
A & =&n,
\end{eqnarray}
and
\begin{eqnarray}
f_{11} & =&\text{tr}_{4}G_{0}(-i\omega_{n},\mathbf{k})\sigma_{z}\otimes%
\sigma_{0}{G}_{0}(-i\omega_{n},\mathbf{k})\sigma_{z}\otimes\sigma_{0}, \\
f_{12} & =&\text{tr}_{4}{G}_{0}(-i\omega_{n},\mathbf{k}){G}_{0}(-i\omega_{n},%
\mathbf{k})\sigma_{z}\otimes\sigma_{0}, \\
f_{14} & =&\text{tr}_{4}{G}_{0}(-i\omega_{n},\mathbf{k})\sigma_{0}\otimes%
\sigma_{y}{G}_{0}(-i\omega_{n},\mathbf{k})\sigma_{z}\otimes\sigma_{0}, \\
f_{15} & =&\text{tr}_{4}{G}_{0}(-i\omega_{n},\mathbf{k})\sigma_{0}\otimes%
\sigma_{x}{G}_{0}(-i\omega_{n},\mathbf{k})\sigma_{z}\otimes\sigma_{0}, \\
f_{22} & =&\text{tr}_{4}{G}_{0}(-i\omega_{n},\mathbf{k}){G}_{0}(-i\omega_{n},%
\mathbf{k}), \\
f_{24} & =&\text{tr}_{4}{G}_{0}(-i\omega_{n},\mathbf{k})\sigma_{0}\otimes%
\sigma_{y}{G}_{0}(-i\omega_{n},\mathbf{k}), \\
f_{25} & =&\text{tr}_{4}{G}_{0}(-i\omega_{n},\mathbf{k})\sigma_{0}\otimes%
\sigma_{x}{G}_{0}(-i\omega_{n},\mathbf{k}), \\
f_{44} & =&\text{tr}_{4}{G}_{0}(-i\omega_{n},\mathbf{k})\sigma_{0}\otimes%
\sigma_{y}{G}_{0}(-i\omega_{n},\mathbf{k})\sigma_{0}\otimes\sigma_{y}, \\
f_{45} & =&\text{tr}_{4}{G}_{0}(-i\omega_{n},\mathbf{k})\sigma_{0}\otimes%
\sigma_{x}{G}_{0}(-i\omega_{n},\mathbf{k})\sigma_{0}\otimes\sigma_{y}, \\
f_{55} & =&\text{tr}_{4}{G}_{0}(-i\omega_{n},\mathbf{k})\sigma_{0}\otimes%
\sigma_{x}{G}_{0}(-i\omega_{n},\mathbf{k})\sigma_{0}\otimes\sigma_{x}.
\end{eqnarray}
Here $G_{0}(-i\omega_{n},\mathbf{k})$ is the Fourier transformation of $%
G_{0}(\mathbf{r},\tau)$ with Matsubara frequency $\omega_{n}=(2n+1)\pi/\beta$%
; $\text{tr}_{4}$ represents the trace calculation for a $4\times4$ matrix.
Note that we calculate the summation over Matsubara frequency numerically
due to the absence of the analytical expression of $G_0$.

Our numerical results show that $J_{xy}=0$ and $J_{\tau x}=0$ while $J_{\tau
y}\neq 0$ for FF superfluids, which is reasonable considering that the
symmetry of the quasi-particle excitations along the $y$ direction is broken
while that along the $x$ direction is kept. In the absence of Zeeman fields,
these parameters can be analytically written as
\begin{eqnarray}
J_{xx} &=&J_{yy} \\
&=&\frac{\hbar ^{2}}{4m}\left[ n-\frac{1}{(2\pi)^2}\int d{\bf k}\sum_{L=\pm
}\left( \frac{\alpha ^{2}m}{4E^{L}}\tanh (\beta E_{L}/2)\left( 1+L\frac{%
\epsilon _{\mathbf{k}}^{2}}{|\alpha |k|\xi _{\mathbf{k}}|}\right) +\frac{%
\beta m}{8\hbar ^{2}}\left( \alpha -L\frac{\hbar \xi _{\mathbf{k}}k}{m|\xi _{%
\mathbf{k}}|}\right) ^{2}\text{sech}(\beta E_{L}/2)^{2}\right) \right] ,
\notag \\
P &=&\frac{1}{8(2\pi)^2}\int d{\bf k}\sum_{L=\pm }\left[ \frac{\beta
}{2}\text{sech}(\beta E_{L}/2)^{2}\left( \frac{\xi _{\mathbf{k}}}{E_{L}}%
\right) ^{2}\left( 1+L\frac{|\alpha |k}{|\xi _{\mathbf{k}}|}\right) ^{2}+%
\frac{1}{E_{L}}\tanh (\beta E_{L}/2)\left( \frac{\Delta_0 }{E_{L}}\right) ^{2}%
\right] , \\
Q &=&n,
\end{eqnarray}%
and $J_{xy}=J_{\tau y}=J_{\tau x}=0$. Here the quasi-particle excitation
spectrum is
\begin{equation}
E_{L}=\epsilon_{\bf k}^2+\alpha ^{2}k^{2}+2Lk|\alpha \xi _{%
\mathbf{k}}|,
\end{equation}%
with $\xi _{\mathbf{k}}=\hbar ^{2}k^{2}/2m-\mu $ and
$\epsilon_{\bf k}=\sqrt{\xi({\bf k})^2+\Delta_0^2}$. These expressions are
exactly the same as previous results~\cite{MingGongPRLS}. For the FF
superfluids, $Q_{y}\neq 0$, leading to $J_{xx}\neq J_{yy}$ corresponding to
anisotropic superfluid densities.

To obtain the low energy excitation spectrum, we write the effective action
in Fourier space
\begin{equation}
S_{\rm eff}^{\rm fluc}=\frac{1}{2}\sum_{\mathbf{k}n}\left(
J_{xx}k_{x}^{2}+J_{yy}k_{y}^{2}-J_{\tau y}E_{\rm sw}(\mathbf{k})k_{y}-PE_{\rm sw}(%
\mathbf{k})^{2}\right) \theta (n,\mathbf{k})\theta (-n,-\mathbf{k}),
\end{equation}%
where $\theta (n,\mathbf{k})$ is the Fourier transformation of $\theta (%
\mathbf{r},\tau )$; $i\omega _{n}$ has been taken analytically to $E_{\rm sw}(%
\mathbf{k}+i0^{+})$. The low energy excitation spectrum can be obtained by
\begin{equation}
PE_{\rm sw}(\mathbf{k})^{2}+J_{\tau y}k_{y}E_{\rm sw}(\mathbf{k}%
)-J_{xx}k_{x}^{2}-J_{yy}k_{y}^{2}=0,
\end{equation}%
which leads to the dispersion
\begin{equation}
E_{\rm sw}(\mathbf{k})=\frac{-J_{\tau y}k_{y}+\sqrt{J_{\tau
y}^{2}k_{y}^{2}+4P(J_{xx}k_{x}^{2}+J_{yy}k_{y}^{2})}}{2P}.
\end{equation}%
Clearly, the dispersion along each direction is linear around $k=0$, with
the slope (i.e. sound speeds) written as
\begin{eqnarray}
v_{x} &=&\sqrt{\frac{J_{xx}}{P}}, \\
v_{y+} &=&\frac{-J_{\tau y}+\sqrt{J_{\tau y}^{2}+4PJ_{yy}}}{2P}, \\
v_{y+} &=&\frac{J_{\tau y}+\sqrt{J_{\tau y}^{2}+4PJ_{yy}}}{2P}.
\end{eqnarray}%
In normal superfluids where $J_{\tau y}=0$ and $J_{xx}=J_{yy}$, the speeds of
sound are isotropic. However, in FF superfluids where $J_{\tau y}\neq 0$ and
$J_{xx}\neq J_{yy}$, they are anisotropic and even different along the
positive and negative \textit{y} directions. It it important to note that
anisotropic sound speeds also happen in anisotropic systems such as equal
Rashba-Dresshaul spin-orbit coupled Fermi gases~\cite{Melo2014arXivS} where $%
J_{xx}\neq J_{yy}$, but in that system $J_{\tau y}=0$, implying that they are
the same along the positive and negative \textit{y} directions.

The integration of spin-wave part yields
\begin{equation}
S_{\rm eff}^{\rm sw}=\sum_{\mathbf{k}}\text{ln}\left(1-e^{-\beta E_{\rm sw}(\mathbf{k}%
)}\right).
\end{equation}
In the presence of spin-wave excitations, the generalized saddle point
equation and particle number equation become $\partial\Omega^0/\partial%
\Delta_0=0$,$\partial\Omega/\partial Q_y=0$, and $\partial\Omega/\partial%
\mu=-n$.

\section{S-2. GENERALIZED Kosterlitz-Thouless RELATION}

For FF superfluids with anisotropic superfluid densities, the KT relation~%
\cite{Bere1971S,ThoulessS} corresponding to isotropic superfluid densities
should be generalized. We consider a free vortex $\theta _{\rm v}=\text{arctan}%
(\rho ^{2}y/x)$ emerging at the temperature $T$ in a FF superfluid, the
free-energy change of the system is
\begin{equation}
F=U_{\rm v}-TS_{\rm v},
\end{equation}%
where $U_{\rm v}$ is the energy of the vortex in a system of size $R$
\begin{eqnarray}
U_{\rm v} &=&\int d\mathbf{r}\left[ J_{xx}(\partial _{x}\theta
_{\rm v})^{2}+J_{yy}(\partial _{y}\theta _{\rm v})^{2}\right]  \\
&=&\pi \sqrt{J_{xx}J_{yy}}\text{ln}(R/a),  \notag
\end{eqnarray}%
with the size of the vortex core $a$. The entropy of the vortex is
\begin{equation}
S_{\rm v}=2\text{ln}(\frac{R}{a}),
\end{equation}%
because the number of configurations that a vortex localizes in a system is $%
(R/a)^{2}$. Here we set $k_{b}=1$. It turns out that a single vortex can be
thermally excited when $F=0$, leading to the BKT temperature
\begin{equation}
T_{\rm BKT}=\frac{\pi }{2}\sqrt{J_{xx}J_{yy}}.
\end{equation}%
This is the generalized KT relation (Eq.(4) in the main text)~\cite{Melo2014arXivS}, which becomes KT
relation when $J_{xx}=J_{yy}$.

\section{S-3. BKT TEMPERATURE, SUPERFLUID DENSITY, AND SOUND SPEEDS}
\begin{figure}[t]
\includegraphics[width=3.4in]{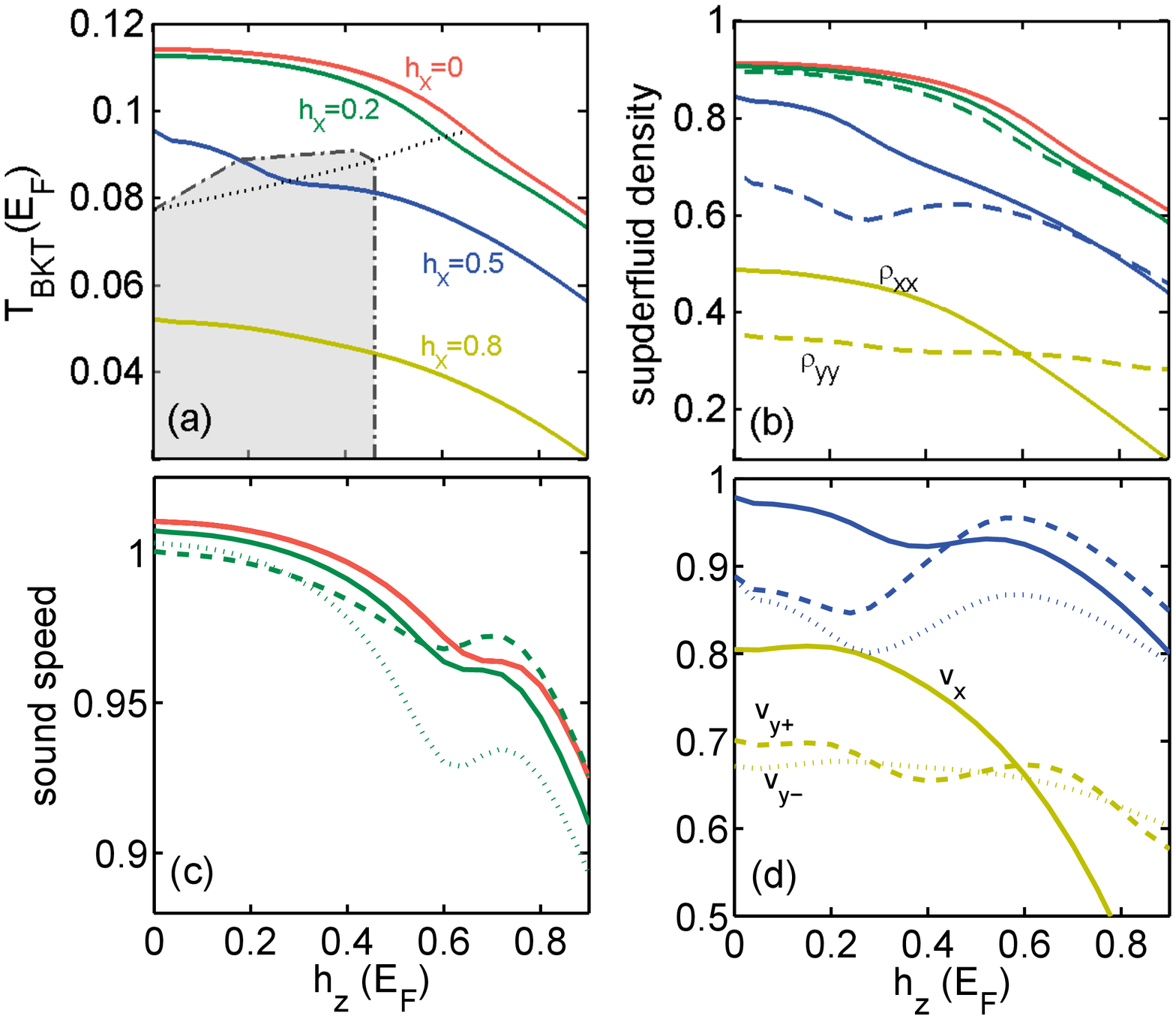} .
\caption{(Color online)  Plot of $T_{\rm BKT}$ (in (a)), superfluid density (in
(b)) and sound speeds (in (c) and (d)) evaluated at $T_{\rm BKT}$, as a function
of $h_{z}$ with $h_{x}$ fixed. All states are FF superfluids except the one
with $h_{x}=0$. The black dotted line connects the topological transition
points as $h_{x}$ varies. The grey area surrounded by the dashed-dotted line
maps out the gapless states region. In (b) the solid and dashed lines
label the superfluid densities along $x$ and $y$ directions respectively
and in (c) and (d) the solid, dashed and dotted lines label the sound
speeds along $x$, positive $y$, and negative $y$ directions respectively.
The red, green, blue, and yellows lines correspond to $h_x=0,0.2E_{\rm F},0.5E_{\rm F},0.8E_{\rm F}$
respectively in all figures.
Here $\protect\alpha %
K_{F}=E_{\rm F}$ and $E_{b}=0.3E_{\rm F}$.}
\label{TS_FF}
\end{figure}
In the main text, we have plotted the BKT temperature, superfluid density,
and sound speeds as a function of out-of-Zeeman field $h_z$ for fixed in-plane
Zeeman field $h_x$ in Fig. 3. Here, we provide more data ($h_x=0.2E_{\rm F}$ and $h_x=0.8E_{\rm F}$) in Fig~\ref{TS_FF}.
The sound speeds are anisotropic and the one along the \textit{x} direction fall
quicker than those along the \textit{y} direction with increasing $h_{z}$ as clearly
shown by the yellow lines in Fig.~\ref{TS_FF}, leading to $v_y>v_x$.
The reason is the symmetry restoration of the Fermi surface at large $h_{z}$,
where $Q_{y}$ begins decreasing, and this leads to the reverse of superfluid densities that
$\rho _{xx}<\rho _{yy}$ at large $h_{z}$.

\section{S-4. VORTEX-ANTIVORTEX VELOCITY FIELD STRUCTURE}
\begin{figure}[t]
\includegraphics[width=3.4in]{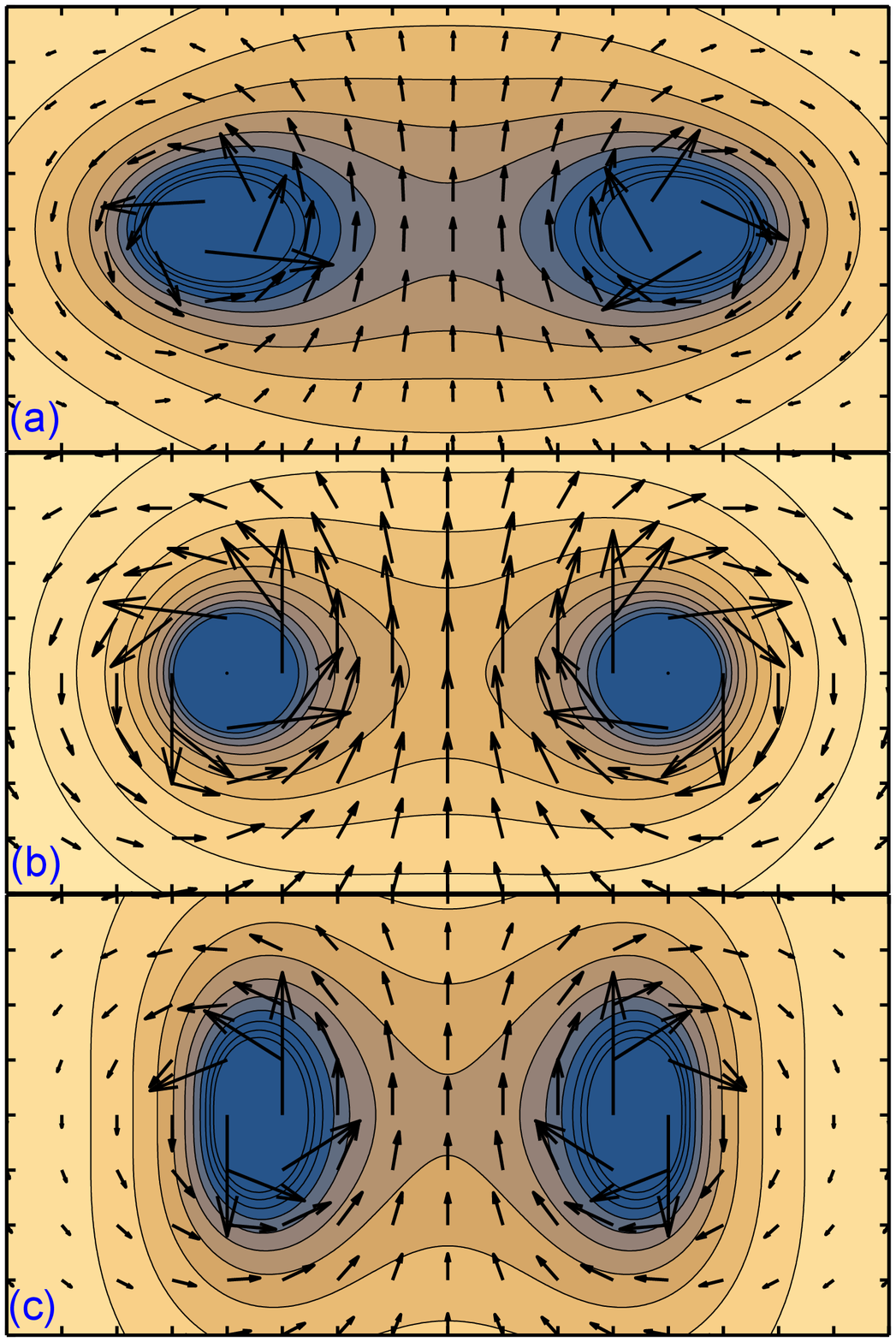} .
\caption{(Color online) Vortex-antivortex structure for topological
superfluids corresponding to $\protect\rho _{xx}>\protect\rho _{yy}$ in (a)
with $h_{z}=0.04E_{\rm F}$, $\protect\rho _{xx}=\protect\rho _{yy}$ in (b) with $%
h_{z}=0.6E_{\rm F}$, and $\protect\rho _{xx}<\protect\rho _{yy}$ in (c) with $%
h_{z}=0.8E_{\rm F}$, evaluated around $T_{\rm BKT}$. Here $\protect\alpha %
K_{\rm F}=E_{\rm F} $, $E_{b}=0.3E_{\rm F}$, and $h_{x}=0.8E_{\rm F}$. }
\label{VAVStruc}
\end{figure}
In the main text, we have shown that anisotropic V-AV pairs emerges across
the BKT temperature. The anisotropy originates from the phase field that
depends on the ratio of superfluid densities along different directions.
Here, to visualize V-AV pairs, we plot their velocity fields in Fig.~\ref%
{VAVStruc}. It clearly shows that there are three distinct vortex cores:
elliptical with the major axis along the \textit{x} direction in (a),
circular in (b), and elliptical with the major axis along the \textit{y}
direction in (c), corresponding to $\rho _{xx}>\rho _{yy}$, $\rho _{xx}=\rho
_{yy}$, and $\rho _{xx}<\rho _{yy}$ respectively. We note that although V-AV
pairs have circular structure in (b), the same as traditional BCS
superfluids, the superfluids have Cooper pairs with finite center-of-mass
momenta.

\end{widetext}
\end{document}